\documentclass[sigconf]{acmart}
\AtBeginDocument{%
  \providecommand\BibTeX{{%
    \normalfont B\kern-0.5em{\scshape i\kern-0.25em b}\kern-0.8em\TeX}}}

\copyrightyear{2023}
\acmYear{2023}
\setcopyright{acmlicensed}\acmConference[CHI '23]{Proceedings of the 2023 CHI Conference on Human Factors in Computing Systems}{April 23--28, 2023}{Hamburg, Germany}
\acmBooktitle{Proceedings of the 2023 CHI Conference on Human Factors in Computing Systems (CHI '23), April 23--28, 2023, Hamburg, Germany}
\acmPrice{15.00}
\acmDOI{10.1145/3544548.3580866}
\acmISBN{978-1-4503-9421-5/23/04}

\begin{document} 

\title[Dynamics of eye-cursor coordination]{Dynamics of eye-hand coordination are flexibly preserved in eye-cursor coordination during an online, digital, object interaction task}

\author{Jennifer K. Bertrand}
\email{jenniferkbertrand@gmail.com}
\orcid{0000-0002-4503-8059}
\affiliation{%
  \institution{University of Alberta}
  \streetaddress{116 Street & 85 Avenue}
  \city{Edmonton}
  \state{Alberta}
  \country{Canada}
  \postcode{T6G-2R3}
}

\author{Craig S. Chapman}
\email{craig.s.chapman@ualberta.ca}
\orcid{0000-0001-7558-3261}
\affiliation{%
  \institution{University of Alberta}
  \streetaddress{116 Street & 85 Avenue}
  \city{Edmonton}
  \state{Alberta}
  \country{Canada}
  \postcode{T6G-2R3}
}



\begin{abstract}
Do patterns of eye-hand coordination observed during real-world object interactions apply to digital, screen-based object interactions? We adapted a real-world object interaction task (physically transferring cups in sequence about a tabletop) into a two-dimensional screen-based task (dragging-and-dropping circles in sequence with a cursor). We collected gaze (with webcam eye-tracking) and cursor position data from 51 fully-remote, crowd-sourced participants who performed the task on their own computer. We applied real-world time-series data segmentation strategies to resolve the self-paced movement sequence into phases of object interaction and rigorously cleaned the webcam eye-tracking data. In this preliminary investigation, we found that: 1) real-world eye-hand coordination patterns persist and adapt in this digital context, and 2) remote, online, cursor-tracking and webcam eye-tracking are useful tools for capturing visuomotor behaviours during this ecologically-valid human-computer interaction task. We discuss how these findings might inform design principles and further investigations into natural behaviours that persist in digital environments.
\end{abstract}

\ccsdesc[500]{Human-centered computing~Empirical studies in HCI}
\ccsdesc[500]{Human-centered computing~Web-based interaction}
\ccsdesc[300]{Human-centered computing~User centered design}
\ccsdesc[300]{Human-centered computing~Laboratory experiments}

\keywords{eye-tracking, cursor-tracking, quantitative methods, eye-cursor coordination, object interaction}


\maketitle

\section{Introduction}
At the core of all well-considered user experiences is the user themselves. So-called human-centered designs incorporate behaviour, cognition, and perception into their product. An early example is the psychophysical mapping of human sensitivity to flickering light. Over hundreds of years, scientists learned that a light display refreshing at a minimum of 30 Hz appeared continuous to the human eye - setting the benchmark for early computer screens.

Of course, perceiving the world is only part of a user’s experience - they also interact with their environment to achieve their goals. Like the display refresh-rate example, principles of human interaction can dictate good design. Fitts’ Law \cite{fitts_information_1954} is one such finding that is now adopted as a design principle in human-computer interaction (HCI) \cite{seow_information_2005,mackenzie_fitts_1992}. In this seminal work, Fitts quantified the speed-accuracy tradeoff for movements of different amplitudes (how far you need to move) to targets of different sizes \cite{fitts_information_1954}. Put succinctly, he showed a lawful relationship whereby larger amplitude movements and smaller targets both result in longer movement times. These real-world findings have since been explored in depth in an HCI context\cite{mackenzie_fitts_1992}, informing and assessing the design of two- \cite{mackenzie_extending_1992} and three-dimensional \cite{grossman_pointing_2004} pointing devices, the soft (virtual) QWERTY keyboard \cite{mackenzie_text_1999, william_soukoreff_theoretical_1995}, and the properties and placement of interactive web elements \cite{lin_prediction_2020, karousos_effortless_2013, roy_incorporating_2021, mcguffin_fitts_2005}.

Critically, user experience (UX) is best thought of as a dynamic cycle of perception and action whereby the information we need to guide our upcoming actions is informed by where we look; how we move then shapes the environment causing changes in the perceptual experience. Consider, for example, the coordinated effort required of the visual and motor systems to safely pick up a cup full of hot coffee. Before any movement, the eyes will fixate the drink, leading the action of the hand by about 500 milliseconds. Seamlessly, as soon as the mug is grasped, the eyes will move to look at the next object for action, like a sugar packet, well before the hand is finished moving the hot drink \cite{land_roles_1999}. These patterns of visuomotor coordination are ubiquitous and stereotypic, appearing in human and non-human primates \cite{arora_eye-head-hand_2019, ngo_active_2022} alike. Here we ask, in the same way real-world findings have informed computer screen and keyboard design, can principles of real-world eye-hand coordination help inform UX design for digital interactions?

To approach this question, we used an online, eye-tracking-enabled platform (Labvanced; \cite{finger_labvanced_2017}) to create a screen-based version of a real-world object interaction task \cite{lavoie_using_2018}. Instead of moving cups to targets on a table, crowdsourced participants (N = 51) dragged circles to targets on their computer screen while we recorded cursor movements and webcam eye-gaze coordinates. We had two primary motivations - first, to explore if the quality of the webcam gaze data (and subsequent processing procedures) would be sufficient to explore visuomotor coordination in our specific task, and second, to quantify if the patterns of eye-cursor coordination would match principles of eye-hand coordination in the real world. Our findings, while preliminary, reveal that both of these are true for this experiment, offering an introduction to entirely new ways of collecting user experience data and suggesting UX design should further explore and consider the tight and environment-invariant principles of visuomotor coordination during target interaction tasks.

\section{Related Works}
\subsection{Real-World Object Interactions}

The tight coupling, in both space and time, of eye movements and motor actions has been well-documented. While much of this research has involved rigid, paradigmatic tasks, those most relevant to UX leverage technological advances to explore eye-hand coordination during self-paced, natural and realistic interactions. For example, the seminal works of Land et al. and Hayhoe measured where people look when making a pot of tea \cite{land_roles_1999} and preparing a sandwich \cite{hayhoe_vision_2000}. Despite the complexity of these tasks and the lack of experimental structure, the researchers were able to break down the tasks into their constituent subtasks (e.g. reaching for the kettle, removing the lid, etc.) to reveal remarkably consistent patterns of eye-hand coordination \cite{land_what_2001}. 

First, even though there are irrelevant objects throughout the kitchen scenes, the eyes only ever fixate on task-relevant objects \cite{land_what_2001}. Critically, these are not the most salient (as defined by low level visual properties like contrast) objects in the visual field; rather, gaze only lands upon task-relevant objects. Second, gaze behaves serially, always fixed upon the current object of manipulation, leading the hand to that object, and when the manipulation is almost complete, moving on to the next object without return \cite{land_what_2001}. In Land et al.’s tea-making task \cite{land_roles_1999}, aggregated across all 94 distinct sub-tasks, the dynamics of the gaze and hand display a consistent pattern: participants fixate on the object to be manipulated for about half a second prior to the hand’s initial movement towards that object and remain fixated there until leaving for the next target about half a second before the completion of the current manipulation.

This general pattern of the eye leading motor action has been found in many contexts, extending beyond kitchen activities to less obvious forms of naturalistic, visually-guided interactions like walking \cite{patla_how_2003, land_eye_2006}, keyboard typing \cite{butsch_eye_1932}, and music playing \cite{furneaux_effects_1999}. Across these interactions, the exact amount of time the eye leads the hand appears to be at least 500 ms (and up to about 1 second), showing some flexibility for the time \cite{deconinck_relative_2011} or accuracy constraints \cite{rand_effects_2010} of the task, or kinematics required for different task contexts \cite{pelz_coordination_2001, johansson_eyehand_2001}. Lavoie et al. \cite{lavoie_using_2018} took a modern approach to investigations of eye-hand coordination during real-world object interaction by combining state of the art motion capture and mobile eye-tracking. In perfect alignment with Land et al. \cite{land_roles_1999}, Lavoie et al. \cite{lavoie_using_2018} found that all object interactions involved the eyes fixating the object at least 500 ms prior to the start of the interaction. Then, within 600 ms of the interaction start, the eyes would leave to look ahead to the next area for interaction \cite{lavoie_using_2018}. This dominant pattern of eye-hand coordination has proven itself highly consistent across real-world object interactions, but, in service of applying real-world findings to digital domains, what happens when we move towards interactions with digital objects?

\subsection{Lab-based Digital Interactions}
Some visuomotor coordination research trades the complexity of an in-lab kitchen for the control of a computer workstation, treating screen-based, digitally-presented objects as a convenient proxy for real-world objects. Nonetheless, patterns of eye leading hand (or computer cursor or manipulandum) are remarkably consistent in the digital domain. In tracing shapes with a cursor, the eye leads the cursor by 223-295 ms \cite{deng_gazemouse_2016}, and in distractorless visual search there is a 190 ms lead \cite{bieg_eye_2010}.  A related paradigm of tracking an unpredictable object on a screen shows that the eye lags behind the target object by 24 ms, whereas the hand lags behind it by 108 ms \cite{danion_different_2018}. During simple reaches towards on-screen targets, the eyes arrive about 386 ms before the hand \cite{sailer_spatial_2000}, although gaze and hand dynamics are flexible to factors like the target’s visibility before or during the reach \cite{van_donkelaar_eye-hand_2000}. Adaptive gaze behaviour is also shown for two-target sequential reaches: the eye anchors to the first target for an extra 95 ms to ensure the hand’s arrival before continuing to a second target \cite{rand_effects_2010}. For visually-guided sequential movements of a manipulandum-handle’s contact with 5 virtual target objects, the gaze arrives at targets 208 ms before contact and leaves 106 ms after contact \cite{bowman_eyehand_2009}. Finally, dragging virtual objects about in a pseudo-touchscreen context also elicits gaze-leading-hand patterns \cite{sims_adaptive_2011}. Taken together, this line of research predominantly studies target-directed reaching and following, but not interaction. Even so, in all contexts, the eyes lead the hand (or cursor) by a few hundred milliseconds.

\subsection{UX-Focused Digital Interactions}
What do eye-cursor coordination patterns look like when digital object interactions are also realistic and ecologically-valid? The HCI domain offers us some answers, however object interactions beyond simple target clicks remain almost entirely unexplored. Early eye-cursor studies employed search engine results pages (SERPs; \cite{rodden_exploring_2007,guo_ready_2010,navalpakkam_measurement_2013}) but mostly considered visuomotor coordination only from a spatial context, measuring the pixel distance between the cursor and eye. Huang et al. \cite{huang_user_2012} did consider timing, finding that the gaze led the cursor by at least 250-, and typically 700 ms during SERP browsing. Of course, with less experimental control, scientists are more likely to find a range of behaviours. Indeed, Smith and colleagues \cite{smith_hand_2000} looked at cursor-pointing to graphical user interface targets and observed at least three eye-cursor coordination patterns: “eye gaze following the cursor to the target”, “eye gaze leading the cursor to the target”, and less commonly, “eye gaze switching between the cursor and target until the target is reached”. The use of multiple strategies during digital yet ecologically-valid tasks is perhaps best illustrated by Liebling and Dumais \cite{liebling_gaze_2014}. Here subjects performed their regular work duties on their office desktop while their gaze and cursor movements were recorded. To begin making sense of the data, the researchers anchored their analysis to cursor clicks, with 32 classes of click-targets determined by metadata records. This rich dataset revealed nuanced coordination - in general, the gaze arrived near the click point 100-200 ms earlier than the cursor. However, the occurrence of or need for coordination varied by target-type: the gaze led the cursor $\sim$30\% of the time for ‘TitleBar’ clicks, yet ‘List’ clicks were gaze-first more than 85\% of the time \cite{liebling_gaze_2014}. In these unconstrained tasks we find additional evidence to support that the eyes lead the cursor, but also see implications for the limitation of this approach when faced with real digital interaction complexity. In the current study, we attempt to strike a balance between ecological validity and experimental control, focusing on a prescribed digital interaction sequence without imposing any constraints on how (e.g. where to look, how fast to move etc.) the sequence should be completed. 

\subsection{Webcam Eye-tracking}
One limitation of almost all of the aforementioned studies is that they occur in the lab. A principle of human-centered design is not only to focus on the user but also to consider the environment and context in which their experience is happening (ISO 9241‑210:2010). One recent technological advancement that might make it possible to study visuomotor coordination in more authentic environments (e.g. users in their own homes on devices they regularly use) is to use webcam data to derive estimates of screen-based gaze behaviour. However, webcam eye-tracking has struggled to establish its utility as a research tool. The reasons are numerous: webcam eye-tracking has a much slower sampling rate ($\sim$10 Hz compared with 100+ Hz in lab \cite{banki_comparing_2022, semmelmann_online_2018, gagne_how_2023}), many users are unable to participate due to a slow internet connection or insufficient hardware \cite{banki_comparing_2022,gagne_how_2023}, uncontrolled lighting can significantly decrease data quality \cite{semmelmann_online_2018, yang_webcam-based_2021, fraser_automated_2021} and, even with optimal conditions, extensive time must be spent calibrating the system (up to 50\% of the study duration as in \cite{semmelmann_online_2018}). Despite these limitations, recent advances, especially in using machine learning to predict gaze location (e.g. Labvanced v2 High Sampling Mode eye-tracking; \cite{finger_labvanced_2017}), offer a path forward, especially where spatial accuracy is the most important feature of the data \cite{semmelmann_online_2018, wisiecka_comparison_2022}. Further, researchers have shown that focusing on fixations to the most relevant areas (i.e. areas of interest, or AOIs) is a reasonable approach for eye-tracking data \cite{holmqvist_eye_2011}. Therefore, a major motivation of the current study was to explore whether state-of-the-art webcam eye-tracking algorithms \cite{finger_labvanced_2017} combined with participation criteria (e.g. processing speed) and AOI-based clustering and analyses would provide sufficient data quality to explore eye-cursor coordination.

\section{Methods} \label{MethodsSection}
\subsection{Participants}
51 adults provided their informed consent to participate in the experiment. Of these, 14 participant datasets were rejected for unsalvageable eye data, and 8 participant datasets were rejected for low trial count (<50\%) after removing trials with procedural or technical errors (see subsection \ref{DataProcess_subsection} - Data Processing and Supplementary Materials for complete data cleaning procedure).  The remaining 29 participants (12 female, 1 undisclosed gender; Age: \textit{M} = 27.07, \textit{SD} = 10.75) were 26 right-hand users and 3 left-hand users. All participants had no prior knowledge about the experiment or its objective. All experimental proceedings were approved by the University of Alberta’s Research Ethics Board (Pro00087329) and were performed in accordance with relevant guidelines and regulations. All participants were recruited using the online crowdsourcing platform Prolific (www.prolific.co) and were paid for their time (6 GBP per hour, $\sim$\$10 CAD per hour).

\subsection{Materials} \label{Materials_subsection}
All participant data was collected online using Labvanced \cite{finger_labvanced_2017}, a browser-based Javascript experimentation platform. The Labvanced platform offers built-in webcam eye-tracking (Labvanced v2 High Sampling Mode eye-tracking \cite{finger_labvanced_2017}) and can record the position of the cursor across time. It was necessary to impose some minimum requirements to achieve stable data collection: only laptop (n = 19) or desktop (n = 10) computers with an audio output (headphones or speakers); only Mac (n = 3), Windows (n = 26) or Linux (n = 0) operating systems and Chrome browser; a webcam with a minimum resolution of 1280 x 720 pixels; a landscape-oriented screen with a minimum of 600 x 600 pixels (\textit{Mode} = 1920 x 1080 px) and a system (including internet connection) capable of collecting at least 10 samples per second of the head’s position for optimal eye-tracking precision (\textit{M} = 14.5 Hz, \textit{SD} = 4.3 Hz).

\subsection{Digital Task Layout} \label{DigitalTaskLayout_subsection}
\begin{figure*}
    \centering
    \includegraphics[width=\linewidth]{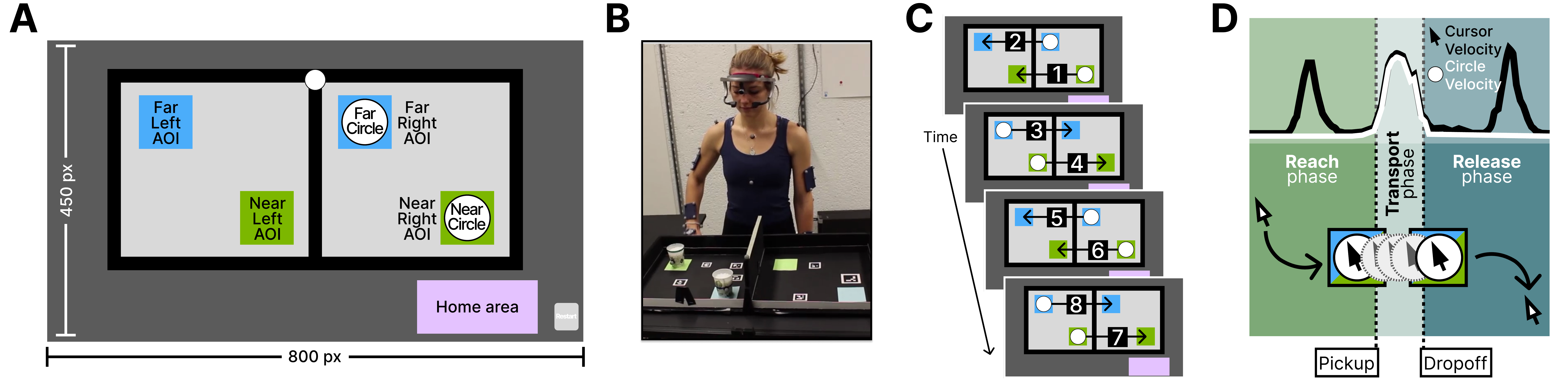}
    \caption{A) The digital task layout. The full screen (800 x 450 pixels) is depicted, where a purple Home area and a Restart button are located in the bottom right corner outside the black-bordered task area. The target Areas of Interest (AOIs) are split into those at the top (blue, Far Left and Right) and those at the bottom (green, Near Left and Right). Participants dragged-and-dropped circles (white, one Near, one Far) between target AOIs. B) A still image of the real-world Cup-Transfer Task we based our digital object interaction task on (from \cite{lavoie_using_2018}). C) The task sequence. A trial involved 8 Moves, with the order and direction of circle movements shown with numbered arrows. The cursor started in the Home area to begin the trial and returned to the Home area after Moves 2, 4, 6, and 8. D) The segmentation of a single object interaction (i.e. one Move) into its events and phases. The top of the panel shows the velocity traces of the circle (white) and cursor (black). The onset of the circle movement was detected as a combination of these velocities and the distance between the cursor and the relevant target AOIs (bottom panel). Together, this defined the Pick-up and Drop-off events, which in turn defined the Reach (green), Transport (light blue) and Release (dark blue) phases.}
    \Description{The figure is separated into 4 parts, with parts A and B at the top of the figure, and C and D at the bottom half of the figure. Part A shows the layout of the screen as a single rectangular frame, as described in the figure caption. The Home area, circles, and AOIs all have text labels overlaid, where the circle at the 'top' of the screen is the Far circle and the circle at the 'bottom' of the screen is the Near circle. The AOIs are presented in a staggered manner, where the Far AOIs are aligned in their both being at the same height on the screen, but are both more leftward than the Near AOIs. The space between the Far AOIs as a pair and the Near AOIs as a pair is the same, they are just staggered in the horizontal plane of each other. The pixel sizing of the screen is shown exterior to the screen image, showing 800 pixels in length and 450 pixels in height. Part B shows a photographed image of the real-world task, where a person is standing in front of a table-top and is glancing downwards at the tabletop. They are wearing a head-mounted eye-tracker, and have many motion-capture markers and plates on their bare arms. In front of them is the real-world apparatus used, which is a 3D version of the screen described in Part A, but instead of circles, two small cups can be seen. Below the cups are the AOIs in the same configuration as described in Part A. Part C depicts the sequence of movements in the task, and shows the 8 movements split into 4 screens, 2 movements per screen. The movements are numbered with arrows showing the location the circle is to be moved. The sequence is described in full in the main text. Part D is an illustration of the segmentation approach we take. The figure is a tall rectangular shape, and is divided into three tall sections by color. The left section is labeled Reach phase, the middle, skinniest section is labeled Transport phase and the right section is the Release phase. The boundaries on each side of the middle Transport phase are dotted, with the left boundary labeled Pick-up and the right boundary labeled Drop-off. Two velocity profiles are depicted at the top of the figure that span the three sections. One shows circle velocity and one shows cursor velocity. The notable information depicted here is the circle velocity rising from zero to a peak and back down again only once, during Transport, while the cursor velocity has a peak in each phase. The bottom of the figure shows an illustration of the mouse approaching a circle (in an AOI) during the Reach phase, the implied movement of the circle during the Transport phase, and the mouse moving away from the circle in the Release phase.}
    \label{fig:Fig1}
\end{figure*}

We modeled our digital task layout (see Figure \ref{fig:Fig1}A) to mirror Lavoie et al.’s Cups Task apparatus \cite{lavoie_using_2018} (see Figure \ref{fig:Fig1}B), which featured a short-walled table-top surface with a midline partition, two cups, 4 AOIs, a Home area, and a fixation sphere. In the real world, participants stood next to the counter-height apparatus, looking down on the surface. Thus, we designed our screen-based version to appear like a flat, bird’s eye view of the real-world task. All real-world elements were proportionally scaled to a 800 x 450 pixel frame in Labvanced (and later, automatically scaled by Labvanced to each participant's screen resolution). Like the real-world version, the Far Left and Right AOIs (FLAOI/FRAOI) were colored blue, the Near Left and Right AOIs (NLAOI/NRAOI) were green, and the Home area was purple. The real-world white paper cups filled with white beads were modeled as white circles. As a proxy for haptic feedback, we designed the circles and Home area to be responsive to cursor hover by darkening in color whenever the cursor landed within their borders. Finally, select elements were introduced to the digital task space in service of loosely replicating the real-world, experimenter-guided experience: a restart button was always available should the participant realize they made a movement sequence error, text appeared with instructions if the participant took more than three seconds to move into their starting position, and a highlighted border appeared around the Home area to mark the important start and end events of the trial.

\subsection{Task}
Our task was an adaptation and extension of an established object interaction task from the real world \cite{lavoie_using_2018}. We doubled the number of object interactions within a sequence, and transformed the task to a digital, screen-based version. Critically, Lavoie’s real-world task was designed to be segmented in time and space to allow for an examination of eye-hand coordination measures around the critical Pick-up and Drop-off interaction events \cite{lavoie_using_2018}. By adopting a similar structure, our analysis also relies on segmentation centered on these key time points. Our 8-movement sequence (see Figure \ref{fig:Fig1}C) is as follows: with the cursor always beginning at Home, Move 1 was a Pick-up of the Near circle at NRAOI, with its Drop-off at NLAOI. Immediately after, the Far circle was picked up from FRAOI and was transported to its Drop-off at FLAOI (Move 2). The cursor then returned to the Home position (as was required after every two object interactions). Moves 3 and 4 were a reflection of the first movements: the Far circle was picked up from FLAOI and moved back to FRAOI, and the Near Circle was picked up from NLAOI and moved back to NRAOI. After the cursor returned to Home, the Far circle was transported from FRAOI to FLAOI (Move 5), and then the Near circle moved from NRAOI to NLAOI (Move 6). This pattern was again reflected after the cursor visited Home, with Move 7 the pickup of the Near circle from NRAOI to drop off at NLAOI, and Move 8 the pickup of the Far circle from FLAOI to FRAOI. The cursor returned to Home to end the trial.

\subsection{Procedure}
Prolific (www.prolific.co) participants were provided a study link and a detailed study description that included an estimate of the study’s duration (1 hour), the hardware requirements, and instructions for avoiding technical complications (included in Supplementary Materials). Clicking the study link launched the full-screen Labvanced window and requested webcam device permission. Participants failing to meet the minimum requirements would receive an error or warning message immediately. Barring no issues, participants would first read a consent form and provided they gave their informed consent, would then answer a brief demographic and hardware survey.

Next, participants would proceed self-paced through the task instructions. Following online research best practices (e.g. \cite{gagne_how_2023}), we developed extensive instructions including an instructional video (see Supplementary Materials), and gave task directions in a way that required participant engagement. Participants were informed that the task was a screen-based version of a real-world task. They were shown a picture of the real-world task (see Figure \ref{fig:Fig1}B) and told that the circles in their task were to be thought of as two-dimensional cups. Finally, participants were encouraged to use favourable lighting conditions and were instructed about the use of a virtual chinrest feature, strategies previously shown to improve webcam eye-tracking data quality \cite{semmelmann_online_2018}.

A 5 minute Labvanced eye-tracking calibration followed the instructions, and participants were required to repeat the calibration if the predicted gaze error exceeded 7\% of the screen size.  Lastly, participants completed one guided (click-through) practice trial, and then a second unguided practice trial with time-delayed hints (i.e. only shown if participant paused for three seconds or longer). Participants could repeat the unguided practice trial as many times as they wanted to ensure they understood the prescribed sequence of movements (1.2 unguided practice trials completed on average).

Participants would then complete the 50 self-paced experimental trials. Every 10 trials, they would receive an update about how many trials they had completed. A brief, 7-point eye-tracking re-calibration was performed every 5 trials, enabling the use of Labvanced’s adaptive drift correction feature. After completing the 50 trials, participants were offered a long-form text input field to provide study feedback (if any) and thanked for their time. The study then concluded, with the browser exiting fullscreen mode, and participants receiving compensation via Prolific.

The entire experimental procedure, as a Labvanced study, can be accessed via the link in Supplementary Materials.

\subsection{Data Processing} \label{DataProcess_subsection}
Employing webcam eye-tracking during an online, self-paced, sequential object-interaction task proved to be challenging. The resulting raw data required a number of quality assessments and treatments to ensure its utility for analysis. While this paper centers on our empirical findings, the corresponding methodological contribution of this work is not trivial, and we provide a detailed account of our data processing pipeline in Supplementary Materials. 

\begin{figure}
    \centering
    \includegraphics[width=\linewidth]{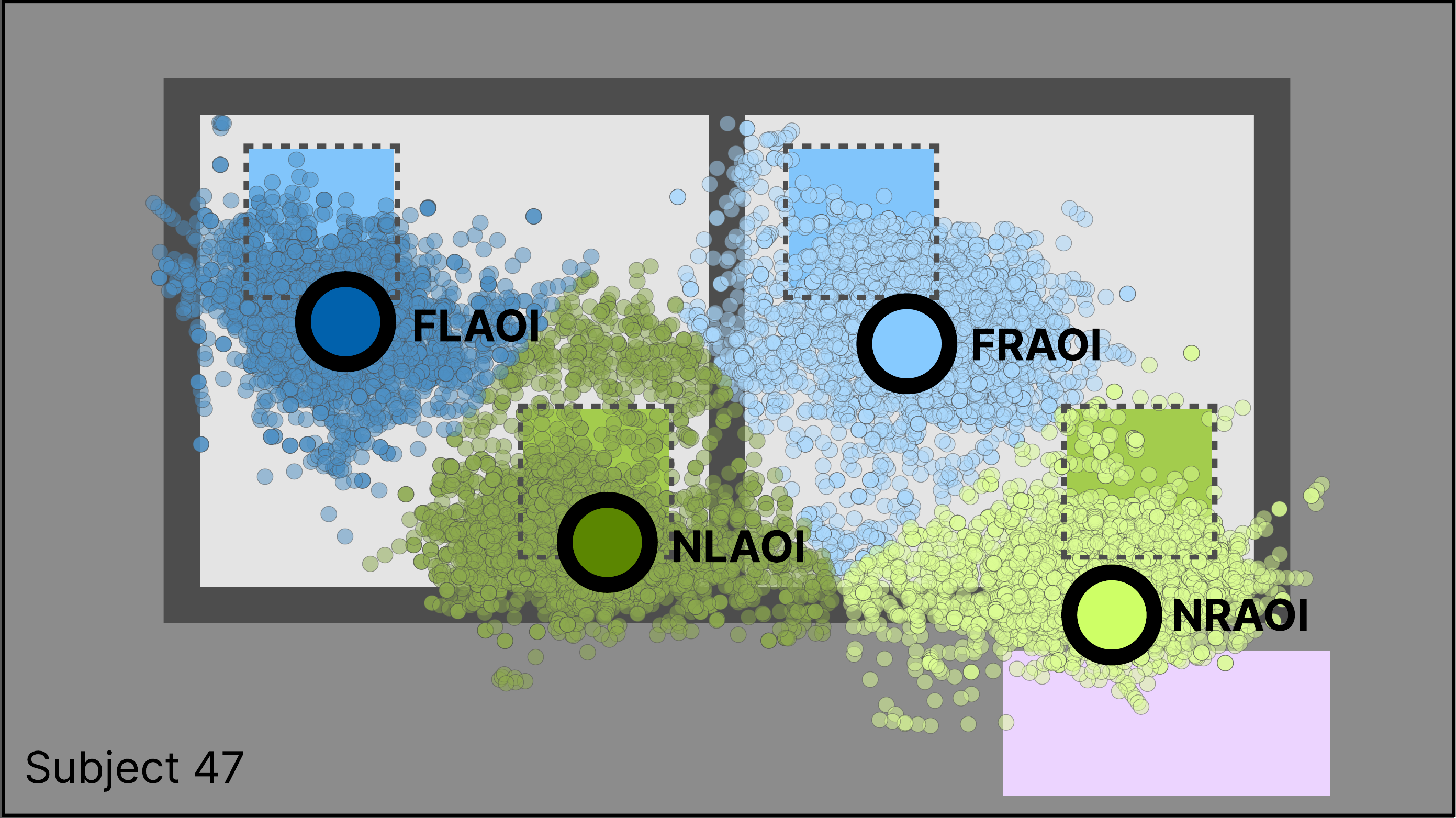}
    \caption{Mapping of one representative participant’s raw eye-data to target AOIs. Raw eye-data samples (small circles) were clustered into 4 bins and, based on their spatial configuration, bins were assigned to one of the 4 target AOIs (dashed boxes). Individual samples are color-coded by their assigned bin: light green -> NRAOI, dark green -> NLAOI, light blue -> FRAOI and dark blue -> FLAOI. Raw-data bin-cluster centroids are shown in corresponding colors as filled circles with black borders. The raw data clearly groups into four clusters but the cluster centroids do not align with the actual task space (skewed down). Our analysis therefore relies on when the eye-data was within each cluster, not the actual space it occupied.}
    \Description{The figure shows a mock-up of the screen, with many tiny circles overlaid, and 4 bigger circles overlaid on the tiny circles. The 4 bigger circles are labeled with the 4 AOI names, with each one siting just below the screen's AOI. The 4 bigger AOI circles are colored in 4 different colors, and the tiny circles underneath and around each AOI circle share the coloring of the AOI to which they are assigned based on the clustering process as described in text and in the figure caption. In general, the coloring indicates distinct clusters of tiny circles, and the purpose is to illustrate the process of clustering the gaze data.}
    \label{fig:Fig2}
\end{figure}

The uncontrolled nature of the online testing environment could give rise to less accurate or spurious gaze predictions. We determined, in a cursory visual inspection of pilot data, that unlike cursor data, gaze data were prone to spatial distortions. That is, while much of the structure of the screen layout of the task was evident in most participants’ gaze data (e.g. many fixations following a pattern shaped like the distribution of targets) these fixation “hot spots” would not necessarily project to the actual target locations - instead they were often shifted or skewed (see Figure \ref{fig:Fig2} and Supplementary Materials for examples). However, if one is primarily interested in which object a person is fixating and when, the exact location of that fixation is mostly irrelevant, and instead you can define and analyze looks to AOIs in relative space. Taking advantage of the fact that our key analyses related to 4 distinct, spatially distributed locations for Pick-up and Drop-off events, we used a data-driven AOI approach. Here, we assumed that participants’ gaze would primarily be driven toward the 4 target locations (NRAOI, NLAOI, FRAOI and FLAOI). Using data from all 50 trials, from all times when the participant had clicked and held the cursor button down, we employed a k-means clustering approach to spatially bin the gaze data into 4 corresponding AOIs (see Figure \ref{fig:Fig2} for a representative participant’s clustering centroids and see Supplementary Materials for additional examples). Thus, our eye-tracking data, while retaining its temporal resolution, was spatially transformed from the 800 x 450 Labvanced coordinate frame into four mutually-exclusive bins: NRAOI, NLAOI, FRAOI and FLAOI. Fourteen of the original 51 subject datasets were rejected because the clustering centroids did not follow the spatial configuration of the AOIs (left targets to the left of right targets, near targets lower on the screen than far targets), suggesting raw gaze data errors beyond a spatial distortion that we could not account for. In the Supplementary Materials, we include a probability density analysis of the accepted clusters in transformed space where we fit bivariate normal distributions to each cluster for each participant and demonstrate that, on average, 28.97\% of eye-tracking data, if linearly transformed to the Labvanced coordinate frame, would fall within the 80 x 80 pixel AOI it was cluster-assigned to. Importantly, only 0.1\% of the eye data risked assignment to any of the 3 non-assigned AOIs in transformed pixel space.

Notably, this approach is not without its risks or limitations. First, by using the data to define the AOIs used for analysis, we run the risk of circularity. Therefore our first test was to ensure that the distribution of looks to each AOI across time matched the time-varying demands of the task. Since our clustering was collapsed across time, this would ensure that the reported looking behaviour was sensitive to the actual task being performed. Second, as discussed in Sections \ref{ResultsSection} - Results and \ref{LimitationsSection} - Limitations and Future Directions, by only creating four cluster-based AOIs, we lose the ability to detect looks to other areas of the display (e.g. the Home or Fixation targets). Since these other areas were not relevant for the majority of our task, and entirely irrelevant to the key interaction events, this was a trade-off we felt was worth making despite the consequence of limiting our approach's applicability to other task designs or research questions. Third, the spatial accuracy assessment of our clustering approach (in Supplementary Materials) highlights the highly effective discrimination \textit{between} AOIs by having effectively no chance of a mis-classified eye-gaze, but it also exposes the challenge of noisier eye-data for \textit{within}-AOI discrimination - had we constrained ourselves to transformed eye data that fell within the boundaries of the actual on-screen AOI-objects, we would have lost more than 70\% of the data. In Supplementary Materials, we represent our clusters as independent bivariate normal probability density functions to visualize their clear spatial cohesiveness, but we acknowledge that the AOI-binning approach may have limited use where targets are less spatially distributed or the task space necessitates unpredictable, dynamic, numerous and/or densely organized targets.

Beyond the eye data cleaning, various steps (as outlined in Supplementary Materials) were taken to ensure participants completed the task correctly. While participants could move the circles about as they pleased, a trial was designed to only advance once all the movements were made. Although various real-time checks were performed to preemptively avoid sequence errors, they still occurred and those trials were removed from the analyzed data. Eight further subjects were removed for having a trial count below 50\% after trial rejection for sequence (and other) errors. Therefore, 29 datasets were included in the following analyses.
\subsection{Segmentation}
In order to explore eye-cursor coordination patterns during object interactions, we needed to define and then automatically identify the 8 movements in each trial and the 2 object interactions (Pick-up and Drop-off) within each movement (see Figure \ref{fig:Fig1}D). This first necessitated re-sampling the cursor and eye data to a common sampling frequency of 60 Hz. Most often this meant that the eye data was upsampled while the cursor data was downsampled. Following Lavoie et al.’s approach \cite{lavoie_using_2018}, we considered the object interaction to include the period when the cursor moves toward the object (Reach: onset = cursor approaching Pick-up location + velocity exceed threshold; offset = Transport onset), the period when the cursor drags the object (Transport: onset = object leaving Pick-up location + velocity exceeds threshold; offset = object approaching Drop-off location + velocity drops below threshold) and the period when the cursor moves away from the object (Release: onset = Transport offset; offset = cursor leaving Drop-off location + velocity drops below threshold), as depicted in Figure \ref{fig:Fig1}D. A Pick-up is said to occur at the transition between Reach and Transport while a Drop-off is said to occur at the transition between Transport and Release. We used our custom Gaze and Movement Analysis (GaMA) software in MATLAB to segment our trial data into 8 movements (Moves 1-8) and each movement into the 3 phases (Reach, Transport and Release). In GaMA, spatial and temporal features of the data are used to automatically define key events, like the start of an object Transport (see \cite{williams_hand_2019}). We followed the principles of Lavoie et al. \cite{lavoie_using_2018} and other real-world examples (\cite{williams_hand_2019,valevicius_characterization_2018}) to segment our data, where thresholds were applied to the cursor and circles’ velocity and AOI-proximity as a means to define the onset and offset of the Reach, Transport and Release phases (see Supplementary Materials for additional details and threshold values). 

\subsection{Data Analysis}
\subsubsection{Dependent Measures}
We had two primary motivations: 1) to determine if the quality of webcam eye-tracking data would be sufficient to explore eye-cursor coordination dynamics for our specific 2D UX context and, if so, 2) to explore, in a preliminary way, if the $\sim$500 ms of fixation time around a manual interaction would be preserved in our specific 2D UX context, despite drastic differences in the physics and style of control.

With respect to 1), it is a hallmark of eye-hand coordination during object interaction that, even though not directly instructed, participants look almost exclusively at task relevant targets (namely the object they are going to interact with and the locations where they are going to move it to and from). As such, our first dependent measure examines, for each target location in the task (4 total, see subsection \ref{DigitalTaskLayout_subsection} - Digital Task Layout), the average time spent fixating that location when it was relevant to the current movement (i.e. a Pick-up or Drop-off location) and the average time spent fixating that location when it was irrelevant to the current movement (i.e. one of the two targets on every movement that are not a Pick-up or Drop-off location).

\paragraph{Average Fixation Duration (ms)} 
Across a given trial, the average time spent fixating on one of the AOIs (NRAOI, NLAOI, FRAOI or FLAOI, see Figure \ref{fig:Fig1}A) when that AOI was a relevant location (a Pick-up or Drop-off location for the current movement) or not. Across the 8-movement sequence, each AOI was relevant and irrelevant an equal number of times.

With respect to 2), fixation time around an interaction consists of two values - how long the eyes are on an object prior to interaction (eye-arrival latency) and how long the eyes linger on an object after an interaction is initiated (eye-leaving latency). Our task involves object manipulations with two interaction events, the Pick-up and the Drop-off. Therefore, we examine the eye arrival and eye leaving latencies for both of these events.

Importantly, it has previously been reported that eye arrival and leaving latencies are not absolute (\cite{lavoie_using_2018}), but can flexibly change based on the demands of the task in general and the durations of each constituent movement and phase in specific. Therefore, to test for these possible within-task adaptations, we also examined eye latencies and each movement in terms of the durations of the Reach, Transport and Release phases.

Based on this motivation, we extracted and analyzed the following measures per movement:

\paragraph{Phase Duration (ms)}
The time spent in each phase (Reach, Transport, Release). 

\paragraph{Eye-arrival latency at Pick-up and Drop-off}
Eye-arrival latency (EAL) at Pick-up was defined as the difference between Transport start time and the time of the eye's arrival at the Pick-up location. EAL at Drop-off was defined as the difference between Transport end time and the time of the eye's arrival at the Drop-off location.

\paragraph{Eye-leaving latency at Pick-up and Drop-off}
Eye-leaving latency (ELL) at Pick-up was defined as the difference between Transport start time and the time of the eye leaving the Pick-up location. ELL at Drop-off was defined as the difference between Transport end time and the time of the eye leaving the Drop-off location.

\subsubsection{Statistical Procedure}
Each dependent measure was analyzed in Jamovi (Version 2.2.5; an open-source statistical software) using a two-factor repeated-measure analysis of variance (RMANOVA). If a two-way interaction was revealed from the omnibus RMANOVA, follow-up single-factor RMANOVAs were performed to test the simple main effects of one factor at all levels of the other factor. Significant main effects were explored with all pairwise comparisons. All reported RMANOVA p-values include a Greenhouse-Geisser correction for violations of sphericity, and all follow-up pairwise comparisons are fully reported in Supplementary Materials (with Bonferroni-corrected p-values).

\section{Results} \label{ResultsSection}
\subsection{Online, webcam eye-tracking can be a suitable method for quantifying gaze behaviours}
\begin{figure}
    \centering
    \includegraphics[width=\linewidth]{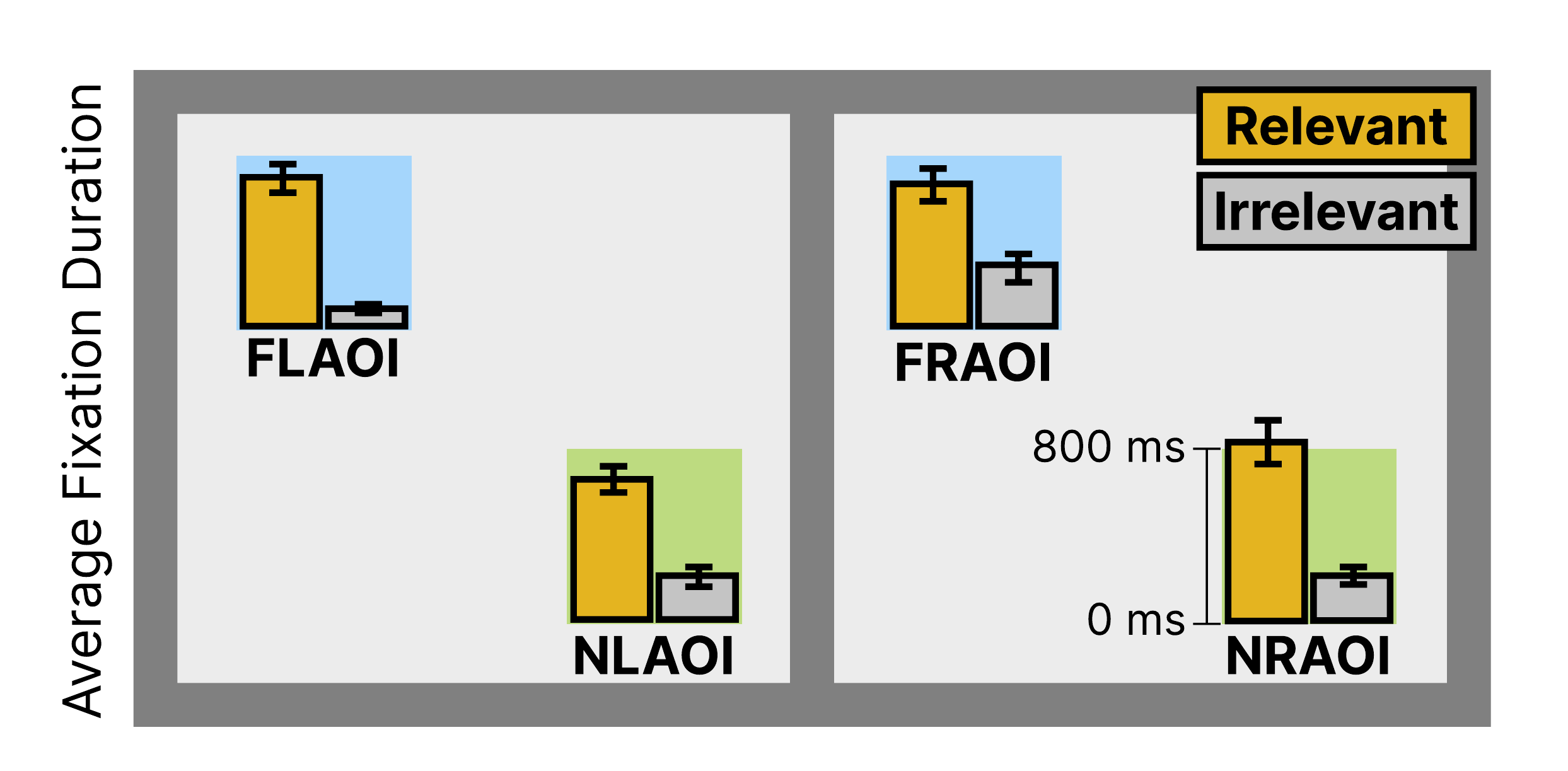}
    \caption{Average fixation duration to each of the four target AOIs (green and blue squares) across movements when that AOI was task-relevant (yellow bars, when that AOI was a Pick-up or Drop-off site for a movement) or task-irrelevant (grey bars). Relevant AOIs were fixated significantly longer than Irrelevant AOIs. Error bars show 95\% confidence intervals of the estimated marginal means.}
    \Description{A mockup of the task space is shown with the 4 AOIs, and overlaid on each AOI is a 2 bar bar-plot. The bars represent the average fixation duration for times when that AOI was relevant or irrelevant, across all 8 movements. The stark result is the much larger bars for relevant as opposed to irrelevant. Error bars are small, and for each AOI, the difference between relevant and irrelevant is very large. The included axis shows all the plots are from 0 to 800 ms, with all the relevant bars around the top end of the axis, and all the irrelevant bars at $\sim$300 ms or less.}
    \label{fig:Fig3}
\end{figure}
As described earlier, our data-driven definition of AOIs leaves us vulnerable to circularity in our analyses. To address this potential criticism, here we look at task relevant timing to check if our approach is valid. Since our AOI clustering is agnostic to timing, any effects of spatial gaze distribution across time that match task demands provide solid evidence for the utility of our approach. Critically, therefore, we show that participants’ gaze fixated more on the Task Relevant AOIs than on the Task Irrelevant AOIs (Figure \ref{fig:Fig3} - see Supplementary Materials for a complementary spatial analysis). These results align favourably with the early, real-world work of Hayhoe and Land \cite{hayhoe_vision_2000, land_what_2001, land_eye_2006} and give credence to our use of webcam eye-tracking as a method for a preliminary investigation of gaze behaviours during our specific online, screen-based, object interaction task. Our 4 x 2 (Position x Task Relevance [Relevant / Irrelevant]) RMANOVA revealed significant main effects of both Position (\textit{F}(1.50, 41.97) = 11.6, \textit{p} < .001) and Task Relevance (\textit{F}(1.00, 28.00) = 329.2, \textit{p} < .001), and a significant interaction between the two factors (\textit{F}(1.83,51.26) = 45.7, p < .001). Post-hoc pairwise follow-ups compared Relevant vs Irrelevant fixation durations at each location - for each location it was fixated more when it was Relevant than when it was Irrelevant (all \textit{p}’s < .001).

This analysis also allowed us to explore for any specific spatial biases in the eye-tracking data recorded in this task. In general, looks to targets were relatively evenly distributed, except for times when looks to other objects in the environment were mis-classified to spatially-proximal target locations. Specifically, at some times, looks to the Home position may have been categorized as looks to the NRAOI and looks to the Fixation position may have been categorized as looks to the FRAOI. This pattern is also visible in the complementary spatial analysis in Supplementary Materials, where the bivariate normal probability density function of the NRAOI shows more dispersion, likely as a result of also capturing some Home position looks. Along the same lines, in this supplemental analysis the FLAOI has the least dispersion, matching its status as the most isolated task-relevant object. As mentioned, however, the overall lack of spatial specificity is a consequence of our AOI clustering but does not appear to add significant noise to our analyses. For a complete analysis of these spatial biases, see Supplementary Materials.

In general, while accounting for the inherent limitations in the design of our study, this analysis offers a demonstration of the sensitivity of webcam eye-tracking. The stark differences in looking time driven by the expected pattern of task relevance, complemented by the low risk of assigning gazes to inaccurate clusters (as evidenced in Supplementary Materials), gave us sufficient confidence to further explore the dynamics of eye-cursor coordination.

\subsection{Digital object interactions yield unique, context-specific movement dynamics}
\begin{figure*}
    \centering
    \includegraphics[width=\linewidth]{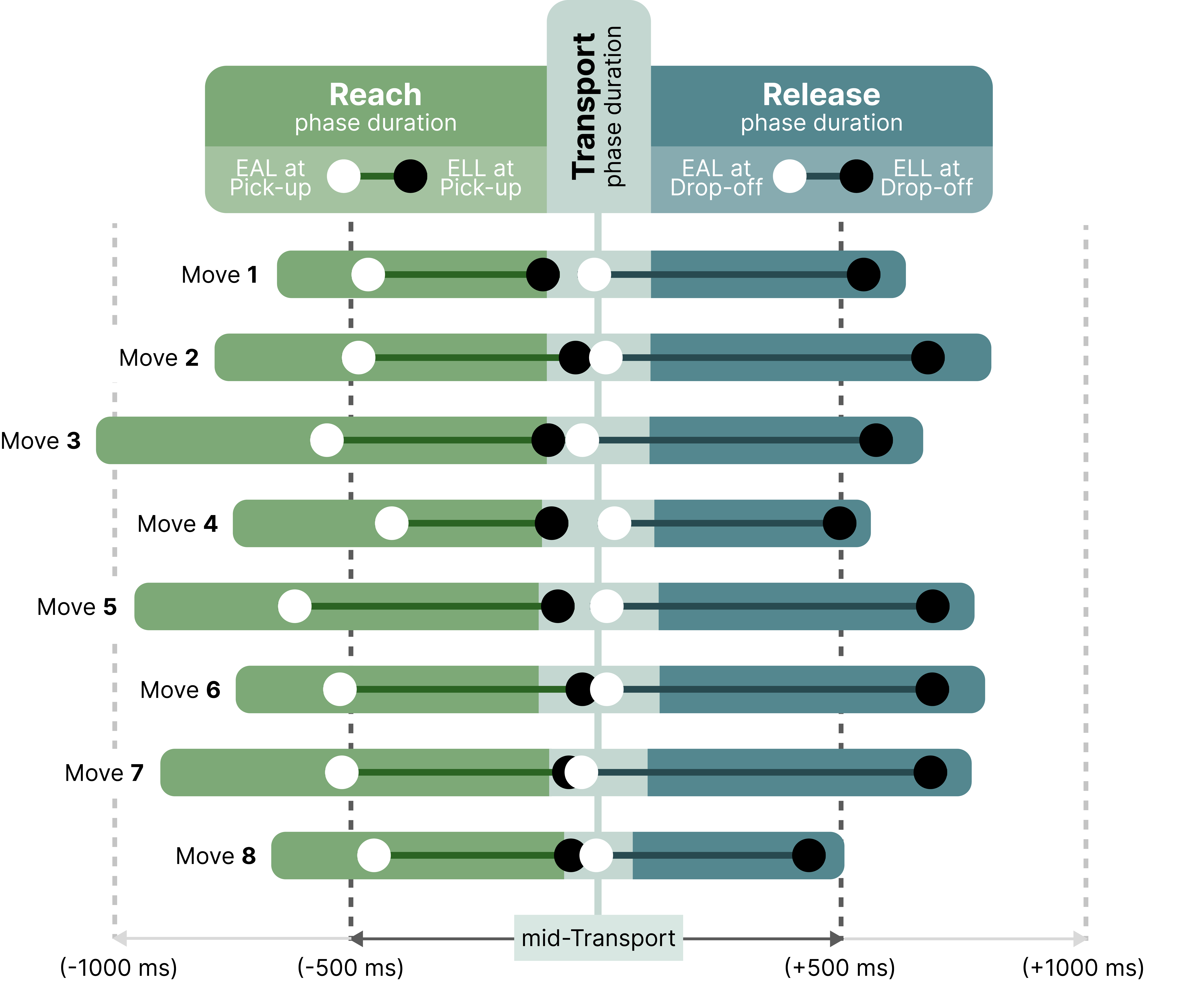}
    \caption{Interaction Phase durations and Eye Latencies at Pick-up and Drop-off across the 8 movements. Each movement is shown as a horizontal bar centered on the middle of the Transport (light blue) phase. Movements start with a Reach (green) as the cursor moves toward the target circle, transition to a Transport as the circle is moved, and end with a Release (dark blue) as the cursor moves away from the circle. The start of Transport is the Pick-up event and the end of Transport is the Drop-off event. White circles show the time the eye arrives (EAL) before each event while black circles show the time the eye leaves (ELL) following an event. Dark green lines connect the Pick-up EAL and ELL while Dark blue lines connect the Drop-off EAL and ELL. Despite the short Transport phases, the eye latencies adapt to ensure at least 500 ms of fixation time around each interaction event.}
    \Description{Horizontal stacked bar plot of the task's interactions, with every move a separate bar. X axis shows time, where all bars are anchored to in the center to the middle of the Transport phase. Reach, Transport and Grasp phases are shown per move, with all moves showing similar short Transports in the center, and Reach and Release phases extending from the left and right ends of the Transport phase, respectively. These show that Reach phases are all at least -500 ms before mid-Transport, and also show Move 3, 5, and 7 extending the furthest left, away from the centre, at or approaching -1000 ms prior to Transport. On the most right end of the horizontal stacks, Release phases all extend to about 500-750 ms after mid-Transport, with slight variations across moves (Move 8 and Move 4 appear the shortest). Overlaid on these phases are the EAL/ELL latencies, where 4 values are shown by 4 circles, coded for EAL or ELL, where EAL+ELL for Pick-up is on the left side of Transport, and EAL+ELL for Drop-off is on the right side of Transport, later in time. EAL at Pick-up consistently hovers around the -500 ms mark, meaning the eyes arrive to the Pick-up location about 500 ms before Pick-up, and the ELL circles sit right inside the Transport phase, meaning the eyes leave right after Transport has begun. For the EAL+ELL at Drop-off values, the EAL at Drop-off is shown for all moves around mid-Transport, and the ELL is shown near the ends of the Release phase, meaning the eyes stay around for about 500 ms after the Transport has been completed.}
    \label{fig:Fig4}
\end{figure*}

As described above, in order to understand the nuances of eye-cursor coordination it is first essential to map the naturally-occurring variations in task demand as indicated by the time spent in each movement and each phase within that movement (Reach, Transport and Release, Figure \ref{fig:Fig4}).  Thus, we used a 3 x 8 (Phase x Movement) RMANOVA to examine phase duration. Both main effects of Phase (\textit{F}(1.58, 44.20) = 138.5, \textit{p} < .001) and Movement (\textit{F}(3.63, 101.68) = 34.1, \textit{p} < .001) were significant, as was their interaction (\textit{F}(5.38,150.53) = 21.0, \textit{p} < .001). Because we were most interested in learning how changes in each phase might impact eye latencies, we examined how phase values changed across movements. The three follow-up single-factor RMANOVAs (Reach/Transport/Release x Movement) each revealed main effects of Movement (Reach: \textit{F}(3.74, 104.80) = 24.5, \textit{p} < .001; Transport: \textit{F}(3.80, 106.41) = 22.2, \textit{p} < .001; Release: \textit{F}(3.13, 87.55) = 29.7, \textit{p} < .001). These results highlight a general pattern of longer Reach phases for movements covering longer screen distances. That is, movements that directly follow a Home visit (i.e. Moves 1, 3, 5, 7) cover more screen distance and elicit longer Reach phases than those that immediately follow a circle movement (i.e. Moves 2, 4, 6, 8) except for Move 1, which has the shortest Home to AOI distance. This finding aligns with the principles of Fitts’ Law.

The single biggest difference between the movement dynamics in the digital compared to real-world task is the duration of the Transport phases. Cursor click and drag movements are \textit{much} faster (around 200 ms) than their real-world counterparts (over 1000 ms from Lavoie et al \cite{lavoie_using_2018}). Besides a quick final move (\textit{M} = 0.143 secs) and some slight differences between other movements, the Transport phases are relatively similar in duration (\textit{M}’s range from 0.205 to 0.255 secs, see Supplementary Materials for full pairwise analysis). Overall, the relative consistency of the Transport duration across the task again reflects that phase timing is primarily related to movement distance - Transport distance is the same for all movements.

Pairwise comparisons (as reported in Supplementary Materials) between movements for the Release phase also follow a pattern of longer phase durations for movements covering more screen distance (between the movement’s drop-off location and the next destination location), as predicted by Fitts’ Law.
\subsection{Eye-cursor coordination during Pick-up interactions resembles the real world, while Drop-off coordination flexibly conforms to the digital context}
Our final motivation was to understand eye-cursor coordination during digital object interaction by examining the latencies between the eye and cursor arriving (EAL) and leaving (ELL) the Pick-up and Drop-off sites across the 8 Movements (Figure \ref{fig:Fig4}). For each of EAL and ELL we ran an 8 x 2 (Movement x Interaction Site [Pick-up / Drop-off]) RMANOVA.

For EAL there were significant main effects of Movement (\textit{F}(3.43, 96.09) = 14.28, \textit{p} < .001) and Interaction Site (\textit{F}(1, 28) = 251.6, \textit{p} < .001), as well as a significant interaction between the two factors (\textit{F}(3.97, 111.2) = 8.37, \textit{p} < .001). Follow-up simple main effect RMANOVAs compared Pick-up and Drop-off EALs across the 8 movements. Drop-off EALs were remarkably consistent, showing no effect of Movement, demonstrating that gaze consistently arrives at a Drop-off location just over 100 ms before the clicked-and-dragged object. Pick-up EALs did show a significant effect of Movement (\textit{F}(3.87, 108.42) = 27.3, \textit{p} = < .001) which aligns with the duration of the Reach phase in which the Pick-up occurred. That is, for movements with a longer Reach phase (e.g. Movements 3, 5, 7) the eye arrives at the Pick-up location earlier - this kind of within-trial flexibility is also observed in the real world \cite{lavoie_using_2018}. Full reporting of the pairwise comparisons is available in Supplementary Materials.

Together, these EAL findings suggest: 1) similar to real-world interactions, during digital Pick-up interactions the eyes arrive about 400-500 ms before the cursor starts to move the object, and 2) unlike real-world interactions, during digital Drop-off interactions the eyes only arrive about 100-200 ms before the dragged object, reflecting the stark differences in the duration of digital versus physical object Transport.

For ELL there were significant main effects for both Movement (\textit{F}(2.93, 82.01) = 22.7, \textit{p} < .001) and Interaction Site (\textit{F}(1, 28) = 340.2, \textit{p} < .001), and also a significant two-way interaction (\textit{F}(4.62, 129.38) = 15, \textit{p} < .001). Follow-up simple main effect RMANOVAs compared Pick-up and Drop-off ELLs across the 8 movements and both were significant (Pick-up: \textit{F}(2.88, 80.7) = 8.2, \textit{p} < .001; Drop-off: \textit{F}(3.6, 100.91) = 24.3, \textit{p} < .001). Despite statistical differences, the timing of the eye leaving the Pick-up location is quite stable and short, ranging from $\sim$35-135 ms. Where Pick-up ELL does vary, it appears to change as a function both of the length of the upcoming transport and as a push-pull with the preceding eye arrival latencies. As an example, Movement 6 has a relatively long Transport phase and comparatively short preceding EAL - this results in it having the longest Pick-up ELL. Full reporting of the pairwise comparisons is available in Supplementary Materials.

The most surprising result from our study, and what stands out as the biggest difference from real-world eye-hand coordination, is how long our participants spend looking at an object \textit{after} they have dropped it off. Here, ELL at Drop-off exceeds 400 ms in all cases and is often more than 600 ms. This is drastically different from the real-world Drop-off ELLs which only range from 140-250 ms \cite{lavoie_using_2018}. This important finding demonstrates that participants compensate for abbreviated digital Transports by having their eyes remain fixated for longer at the location where the object is dragged to. This prolonged Drop-off ELL also scales with duration of the Release phase, with Movements with longer Release phases also showing the longest ELLs. This relationship suggests that the compensatory prolongation of the Drop-off ELL may in part relate to the planning of the next movement following a Release. Full reporting of the pairwise comparisons is available in Supplementary Materials.

Together, our ELL findings further inform the nuances of eye-cursor coordination in digital interactions: 1) like real-world Pick-ups, the eyes wait until the Pick-up happens then quickly leave, and 2) unlike real-world Drop-offs, the eyes dwell at the Drop-off site well beyond the end of the Transport.

\section{Discussion}
In this preliminary investigation, we show that eye-cursor coordination during a specific form of digital object interaction obeys constraints similar to eye-hand coordination during real-world interactions. Specifically, we find the eye dwells on or near the site of a digital interaction for at least 500 ms, almost identical to the amount of time others report in real-world interactions with physical objects. This initial finding demonstrates the potential utility of webcam eye-tracking collected from online, remote, crowdsourced participants as a tool for capturing rich, meaningful, and ecologically-valid visuomotor data.

Our study was designed to make the comparison of digital to real-world interactions as valid as possible. Thus, we adapted a previously reported real-world task \cite{lavoie_using_2018} where the 500 ms minimum dwell time per interaction had previously been quantified. In our digital adaptation of this real-world cup-transfer task, we asked crowdsourced, online participants to perform 50 trials of an 8-movement drag-and-drop sequence (see Figure \ref{fig:Fig1}). While performing this digital interaction task, participants' cursor and gaze positions on the screen were monitored via the Labvanced experiment platform (\cite{finger_labvanced_2017}) using their own computer webcams. As described throughout this study, there are significant challenges to collecting webcam eye-tracking data, and as such, a major objective for this project was to assess its feasibility as a tool for quantifying patterns of dynamic eye-cursor coordination.

Tentatively, and with the caveat that substantial preprocessing was required, we believe that the eye data quality in this task was sufficient to explore eye-cursor coordination for this specific digital context. First, as reported by other research groups collecting online data (i.e. \cite{semmelmann_looking_2017,semmelmann_online_2018,yang_webcam-based_2021}), we experienced high rates of data exclusion (>40\% of participants were not included in analysis, predominantly due to eye-data quality issues) even though we imposed restrictions on the hardware and internet connection of eligible participants. Second, for the participants who were included in the analysis, the eye data required extensive processing. This included reducing the spatial dimensionality from the (x,y) coordinate frame of the screen to 4 data-driven AOI bins (see Figure \ref{fig:Fig2} for a representative subject), which were then mapped to the 4 task-relevant interaction locations. Given this approach departs from conventional eye-tracking analysis, we confirmed its sensitivity by testing if the distribution of gaze to each of these task-relevant AOIs followed the predictions imposed by task demand. Specifically, we show that participants fixated more on interaction locations that were relevant to the current movement (i.e. the target where you were clicking an object and the target where you were dragging it to) than to locations that were not relevant to that movement (see Figure \ref{fig:Fig3}, and Supplementary Materials for the complementary spatial investigation).

It is important to acknowledge that our successful collection and preliminary validation of webcam eye-tracking is in and of itself a significant contribution. With the uncontrolled nature of online tasks, and a technology that relies on consumer-grade hardware, there are many opportunities for noise or error to prevent successful data collection. We worked hard to minimize dropout due to hardware and internet issues by imposing very clear requirements, stated during crowdsourcing and checked during the study initialization. Then, we spent considerable time developing clear and transparent instructions to assist with participant retention. We provided detailed information about potential privacy concerns as well as video and interactive walk-through demonstrations of the task to promote participant understanding (see Supplementary Materials). As described above, our data was then processed using a k-means clustering technique. While clustering the eye-data was a successful approach for this study, it succeeded in part because our task-relevant AOIs were static and spatially distributed. Importantly, this means this approach will not be as successful or even possible for tasks with dynamic AOIs or AOIs that are close together. We further discuss these and other important limitations in Section \ref{LimitationsSection} - Limitations and Future Directions, below.

Our investigation of gaze distribution (see Figure \ref{fig:Fig3} for a temporal assessment, and Supplementary Materials for a spatial assessment) gave us sufficient confidence in the quality of the eye-tracking data to pursue our primary question of whether or not digital eye-cursor coordination in this task would follow similar patterns to real-world eye-hand coordination. Examining visuomotor coordination in tasks designed to promote natural, self-paced behaviours requires the task first be broken into its constituent interactions and then that those interactions be broken into the discrete phases of interaction. Adopting the segmentation strategy employed by Lavoie et al.\cite{lavoie_using_2018} we identified an important distinction between real and digital object interactions: despite both forms of movement being self-paced, digital objects are transported in about 200 ms (see Figure \ref{fig:Fig4}), 4 to 5 times faster than real objects are moved in the real world. Given these movements, it was impossible that the exact pattern of eye-hand coordination observed in the real world would be preserved during digital interactions. That is, during real-world interactions, when transporting an object between locations, the eye will “look ahead” to the drop-off site about 500 ms before the hand and object arrive. But, as just explained, during digital interactions, the \textit{entire} transport lasts about 200 ms, meaning the eye cannot look ahead in the same way. 

Remarkably, our results suggest that the visuomotor system preserves the overall interaction eye-dwell time of at least 500 ms by flexibly adapting the pattern of fixations. Specifically, we quantified eye-cursor latencies around both the Pick-up (cursor clicked to start dragging) and Drop-off (release of cursor click to stop dragging) events. At each event, we calculated how long the eye was looking at the location prior to the event (eye arrival latency, or EAL) and how long the eye remained looking at the location after the event (eye leaving latency, or ELL). As depicted in Figure \ref{fig:Fig4}, for a digital Pick-up, the pattern of gaze is almost identical to a real-world interaction: the eyes arrive at the location 400-500 ms before and stay for about 100 ms after. Given the speed of the Transport phase, eye latencies during digital Drop-off are significantly different from the real world. The eyes arrive around 100 ms before the event, but surprisingly, linger for 400-500 ms after the digital object has been released. With consideration to the preliminary nature of our investigation, we take this as the most important finding in our study: gaze allocation during a digital, self-paced, drag-and-drop object interaction flexibly adapts to the drastically different mechanics of movement to ensure about 500 ms of visual information is acquired from the beginning and end of each interaction movement. Thus, this perfectly aligns with the take-home message of real-world interactions, but the route by which it is achieved is significantly different.

Given this initial evidence for the persistence of $\sim$500 ms of eye fixation towards objects we interact with across real and digital domains (for at least this specific context), what insights might this offer for UX design principles? First, it suggests that as particular digital experiences are being designed, there may be a fundamental lower limit on the pacing with which interactions can occur while respecting the natural cadence of visuomotor coordination. For example, a drag-and-drop movement will take at least one second if executed at a natural speed. A designer wanting to push these movements to be faster might consider rearranging the target locations to make them spatially contiguous, possibly allowing the Pick-up and Drop-off information to be gathered by a single fixation. On the flip side, a designer requiring particularly precise cursor interactions might consider spatially separating the targets, letting the extra distance provide additional time for targeting fixations to occur within the natural rhythms of the task. In both of these cases, these preliminary findings support the notion that a design achieves maximum efficiency not by being the fastest, but rather, by aligning with the demands of a visuomotor system that evolved to optimally coordinate movements \cite{cisek_evolution_2022} on its own time scale.

A second design principle which is more indirectly revealed via this introductory study is the role that feedback plays in facilitating successful interactions. Unlike in the real world where a person using their body to interact with an object typically receives haptic feedback about the interaction, digital interactions rely almost exclusively on visual feedback for confirmation that an interaction is proceeding successfully. In the current study we attempted to boost the visual cues associated with interaction by changing the visual properties of targets based on the cursor position. But, we believe there is more work to be done exploring how additional modalities could move digital interactions toward their real-world counterparts. For example, adding sound cues relevant to interaction, or even more sensitive and dynamic visual cues on objects that are successfully being interacted with may liberate the eyes to move in an even more natural fashion. Some work in high leverage interactions like laparoscopic surgery \cite{panait_role_2009} has shown the utility of this approach. As real and digital worlds become less siloed and share elements across spaces (e.g. virtual and augmented reality), we predict that these mixed interactions will also obey the $\sim$500 ms of viewing time and so too will benefit from the exploration of multimodal feedback cues.

\section{Limitations and Future Directions} \label{LimitationsSection}
Our goal was to measure more ecologically-valid user experiences from a diverse population of participants using their own digital devices in familiar environments. Thus, we collected eye-tracking data from webcams in remotely recruited participants. This meant we sacrificed some experimental control and introduced more eye-tracking noise, leading to a number of notable limitations. Here we describe some of these limitations and for each, offer a future direction for how to test and improve the study.

First, we provide no formal validation of the accuracy of the webcam eye-tracking system. While we attempted to quantify the functional accuracy in both time (see Figure \ref{fig:Fig3}) and space (see Supplementary Materials) a future approach would be to conduct the same experiment under controlled lab conditions while simultaneously recording both webcam and lab-grade, high-resolution eye-tracking data. Of course, a shift to the lab would also remove some of the environmental confounds of remote participation (lighting, hardware differences etc.) and would therefore provide a best-case measure of the magnitude of accuracy difference between webcam and lab-grade systems. Thus, given the known accuracy reduction in the current study, we urge the reader to take these results as preliminary and interpret them with due caution, leaving formal validation as a future opportunity.

Second, the reduced accuracy of the webcam eye-tracking forced us to define and use four, large, mutually-exclusive AOI bins with a spatial distribution roughly matching the actual targets but susceptible to spatial skewing (see Figure \ref{fig:Fig2} and Supplementary Materials). As a result, we are unable to identify exactly where within a defined AOI the gaze was focused. Our analysis assumes that a look within a particular AOI is actually a look toward the relevant target object - an assumption that follows from real-world tasks where gaze is anchored only to task-relevant objects \cite{land_what_2001,hayhoe_vision_2000}. In the Supplementary Materials we present our attempt to quantify some aspects of this assumption by examining the spatial dispersion of the eye data when transformed into the task space. As reported, this analysis suggests that our AOI-binning was useful (unlikely to result in eye-data being mislabelled) but also highlights the remaining noise (most eye data still falls outside of a task-defined AOI). Therefore, the assumption of the specific timing and location of eye-positions relative to AOI boundaries should also be tested in future work. That is, further study is needed to confirm that this pattern of gaze anchoring extends to digital, screen-based contexts, and is also true in uncontrolled, remote settings. Again, a future study is needed to directly test this assumption by conducting an in-lab experiment comparing webcam to lab-grade eye-tracking. 

Third, and perhaps most importantly, the large-AOI approach detailed above presents a significant limitation in our ability to draw definitive conclusions about how digital interactions relate to real-world interactions. This means the results of our second research question should be treated with particular caution as they are based on unvalidated assumptions. As an example, we are unable to provide any quantifiable proof that the arrival and dwell time of eye data binned into four large screen-based AOIs is equivalent, or even a good proxy for, the arrival and dwell time of a real-world fixation to a real-world object. As mentioned above, the spatial dispersion analysis of eye data clusters presented in the Supplementary Materials provides some quantified context for the general validity of this approach, but, again, a potential future solution would be to conduct a laboratory eye-tracking experiment benchmarked against a gold-standard, high-resolution eye tracker. Moreover, this future study could also explore a range of real-world and digital tasks to make the connections between them more clear. As an example, a more consistent real-world variant of the screen-based drag and drop task we employed here would have participants move and slide an object across a table, rather than lift and place as was done in the previous real-world experiment we used as inspiration \cite{lavoie_using_2018}. 

Finally, the previous point about task variety highlights a limitation of our approach regarding its wider generalizability. To be clear, we only examined a single digital-interaction task that was designed to mimic only a single real-world object interaction task, and as such, our claims of this exposing a general property of the required fixation time for successful interaction ($\sim$500 ms) should be tempered. As with how the body can interact with objects in the real world, during digital interactions there are a multitude of ways people can interact with digital objects. Here we limited ourselves to only one form of interaction (drag-and-drop) and as such, any conclusions we draw may only apply to that particular case. It is possible, or even likely, that other styles of interaction (e.g. point-and-click, swipe, hover) would result in different patterns of gaze behavior. 

Taken together, these limitations highlight the need to interpret these results as exploratory and further validate some of the key assumptions. That being said, while accounting for these limitations, this study does suggest the promising power of remote data collection for ecologically-valid, user-centered, visuomotor research. This study also offers practical and automated methods (albeit task-specific) for extracting gaze patterns from unreliable data. Here, those methods provide sufficient quality to conduct an exploratory investigation of eye-cursor coordination under these more challenging conditions. In doing so, we believe we provide encouraging, though preliminary, evidence for some basic principles of gaze behavior during digital interactions. We think these results are exciting to drive future research focused on both validating the webcam results with in-lab experimentation, and exploring visuomotor coordination across the broader digital interaction space. 

\section{Conclusion}
We believe our study introduces a potential new approach to user testing that accounts for aspects of UX that are rarely considered. Specifically, by adopting fully-remote testing of participants from the comfort of their own home, using their own hardware, we are actually testing users in the environment and context in which they’d typically encounter digital products. Moreover, our unique approach of harvesting gaze and movement data and automatically converting the data into objective metrics of experience gives rise to previously untapped insights. Here, in a preliminary investigation, aided by advances in webcam eye-tracking \cite{finger_labvanced_2017}, we used this new level of insight to compare coordination across real and digital worlds, but this information may be valuable in innumerable contexts. The repertoire of digital interactions extends well beyond clicks and drags to points, swipes, flicks, pinches, taps and any number of other actions. Each of these is likely to be accompanied by a stereotyped pattern of natural visuomotor coordination which, when studied through the lens of gaze and movement behaviour, can help refine design processes. Already the benefits of this approach are being seen in real-world applications where scientists are better able to assess the movements of prosthetic limb users \cite{hebert_quantitative_2019} with the goal of helping those patients achieve more functionality in their activities of daily living.

It turns out, people \textit{don’t} move in mysterious ways. Instead, there are particular strategies for effective interactions that are true across drastically different environments. Like Fitts found for speed accuracy tradeoffs \cite{fitts_information_1954}, which retain their relationship on land \cite{fitts_information_1954}, underwater \cite{kerr_movement_1973}, and in space \cite{newman_memory_1999}, here we report that across real and digital interactions, properties of eye-hand and eye-cursor coordination remain constant. By adapting design principles to align with these invariant properties of human performance we stand to improve the user's experience.



\bibliographystyle{ACM-Reference-Format}
\bibliography{references}




\end{document}